\documentclass[reprint,groupaddress,amsmath,amssymb,aps,pra,showkeys,showpacs]{revtex4-1}
\usepackage{graphicx}
\usepackage{color}

\newcommand{\ket}[1]{\mathinner{|{#1}\rangle}}
\newcommand{\bra}[1]{\mathinner{\langle{#1}|}}

\begin{document}
\title{Quantum entanglement driven by electron-vibrational mode coupling}
\author{F. M. Souza}
\email{fmsouza@ufu.br}
\author{P. A. Oliveira}
\author{L. Sanz}
\affiliation{Instituto de F\'isica, Universidade Federal de Uberl\^andia, 38400-902 Uberl\^andia, MG, Brazil}
\date{\today}

\begin{abstract}
In this work, we provided a proof-of-principle of efficient production of maximally entangled states using charged quantum dots coupled to vibrational modes. The physical system consists of two pairs of quantum dots, each pair with a single
electron able to tunnel between the dots, thus encoding a qubit.
The electrons, initially not coupled, interact with two bosonic vibrational modes.
It is demonstrated that the electron-vibrational mode coupling drives to an effective
electron-electron interaction, which is the main mechanism behind the formation
of maximally quantum entangled electronic states. The effect of this coupling
follows a non-monotonic behavior, which is explained through an effective
hamiltonian which takes into account high order transition processes.
\end{abstract}
\maketitle

\section{Introduction}
\label{sec:intro}

Semiconductor nanoestructures have become a promising scenario for implementation of quantum computation,
as originally proposed in the late 90s~\cite{Loss98,Burkard99}. These devices show high versatility in front of the wide set
of degrees of freedom that can be used to encode a qubit~\cite{Zhang18}. Single qubit operations have been reported in several
physical setups including the electronic spin $1/2$ states~\cite{Yoneda18}, where fast two-qubit gates have been implemented
recently in silicon~\cite{Zajac18,He2019}. Other possibilities for quantum computation in semiconductors include
the single-triplet qubit states of two-electrons in GaAs~\cite{Petta2005,Nichol2017}, the exchange-only qubit with
spin states~\cite{Laird10} and the charge degree of electrons~\cite{Hayashi03,Mi18}, among others.

Quantum dots (QDs) show interesting properties due to the confinement of particles.
From all the possibilities, including optical quantum dots~\cite{Scholl19,Borges16,Borges12}
and electronic spin~\cite{Russ18}, the interest on the physics of charged quantum dots has been increasing,
once they are scalable systems where initialization and readout are possible through a process involving
detection even of a single electron~\cite{Kiyama18,park2002}. In this physical system,
the qubits are defined based on the property of electronic tunneling~\cite{Shinkai07,Shinkai09},
with the single-qubit operations being controlled by this effect, together with the manipulation
of the electronic detuning~\cite{Shinkai07,Shinkai09}. The single-molecule electronics has
been an outstanding issue due to its future implementations feasibility of a cheaper and
faster single-electron transistor~\cite{xiang2016,Ratner05}.

To further increase the functionalities of a qubit with electrons in a quantum dot, it is interesting to check coupling effects to nanomechanical degrees of freedom~\cite{park2000,steele20092,lassagne2009}. This kind of interaction plays a significant role, bringing a wealth of interesting effects, such as quantum-shuttles in QDs~\cite{gorelik1998,armour2002,donarini2005}, local cooling~\cite{kepesidis2016}, phonon-assisted transport in molecular quantum dot junctions~\cite{walter2013,Sowa17}, and Franck-Condon blockade~\cite{leturcq2009}, among others. One possibility is the use of carbon nanotubes (CNT), one of the most successful
new materials in view of their broad set of direct applications~\cite{dresselhaus2001}.
When operated as mechanical ressonators, nanotubes show high quality
factors~\cite{lassagne2009,laird2012,moser2016} being possible, for instance, to excite,
detect and control specifical vibrational modes of a CNT with a current being injected
from a scanning tunneling microscopy (STM) tip into a CNT~\cite{leroy2004}. Also, CNT can be used in the implementation of ultrahigh tunable frequency resonators~\cite{deng2016,Chaste11,sazonova2004}, nanoradios~\cite{jensen2007}, ultrasensitive mass sensors~\cite{chiu2008,jensen2008}, and it has been reported strong coupling regimes between single-electron tunneling and nanomechanical motion on a suspended nanotube, tuned via electrical
gates~\cite{benyamini2014}. Regarding applications in micro and nanoelectronics,
carbon nanotubes present balistic conduction~\cite{laird2015} and Coulomb blockade effect in single and double nanotube based quantum dot devices~\cite{steele2009}. Particularly, it was proposed a mechanically induced two-qubit quantum gate and the
generation of entanglement between electronic spin states in CNT~\cite{wang2015} and showed its potential as ``flying'' qubits for electron spin communications over long distances~\cite{deng2016}.

From the theoretical point of view, one successful model to explore the problem of a two-level system interacting to bosons was proposed by Rabi~\cite{Rabi36,Rabi37}, which can be treated both numerical and analytically~\cite{sissueRabi16,PhysRevLett.105.263603,PhysRevLett.107.100401,Duan15}.
A specific approximation, the Jaynes-Cummings (JC) model, becomes the theoretical support behind several quantum phenomena, including the formation of Schr\"odinger cats and quantum logic gates~\cite{RevModPhys.73.565,RevModPhys.85.1103}. The Rabi and JC model have been used in the context of the qubits coupled with bosons~\cite{Gullans15,Palyi12}, where this type of coupling become responsible by single-qubit operations. Alternatively, the interaction between particles and bosonic fields can result in the formation of polarons~\cite{Landau48,polaronBOOK}. Codifying a qubit as a polaron becomes a challenge, once the electron-phonon interaction is generally a mechanism of decoherence~\cite{Stavrou05,PhysRevB.71.205322,Golovach04}, although it was shown that this type of interaction can be used to build a quantum dot maser~\cite{Gullans15}. In what concerns the problem of quantum correlations emergence between qubits, mediated by bosonic modes, some theoretical works are found in literature~\cite{Royer2017fastandhighfidelity,1751-8121-46-33-335301,Bina14}. In recent experimental works, the generation of quantum correlations is demonstrated considering the coupling of exciton with phonons~\cite{krzywda2016phonon}. Other possibility is the use of the coupling with photons~\cite{delbecq2013photon} to control a two-qubit operation.

In this work, we investigate a system composed of two charge qubits interacting
with each other via electron-vibrational mode coupling.
The main goal of our study is to provide a proof-of-principle
that electrons in quantum dots, coupled to high-frequency bosonic nanoresonators,
can be used to generate maximally entangled states of charge qubits.
The role of the electron-boson interaction is quite different in charge qubits
if compared to the spin two-qubits scenario, once in charged quantum dots,
this interaction preserves the state of a single qubit on the electronic degree of freedom,
while creates or annihilates an excitation in the bosonic space.

The paper is organized as follows. In Sec.~\ref{sec:model}, by using the unitary transformation of Lang-Firsov,
we demonstrate that the electron-vibrational mode coupling is responsible for the occurrence of an
effective electron-electron interaction. Then, we demonstrate how to encode two electronic qubits in our physical system. Section~\ref{sec:eigenproblem}
is devoted to the exploration of the signatures of this effective interaction
and correlated phenomena on the spectrum and eigenstates of the model.
In Sec.~\ref{sec:dynamics}, using as reference our previous work on quantum
dynamics on coupled quantum dots~\cite{Oliveira15,souza2017}, we study the formation
of maximally entangled electronic states under specific conditions. The feasibility and robustness against charge dephasing, the main decoherence process in the physical system of our proposal, is discussed in Sec.~\ref{sec:feasibilty}. Section~\ref{sec:summary} contains our final remarks.

\section{Model}
\label{sec:model}
\begin{figure}[tb]
\centering\includegraphics[width=1\linewidth]{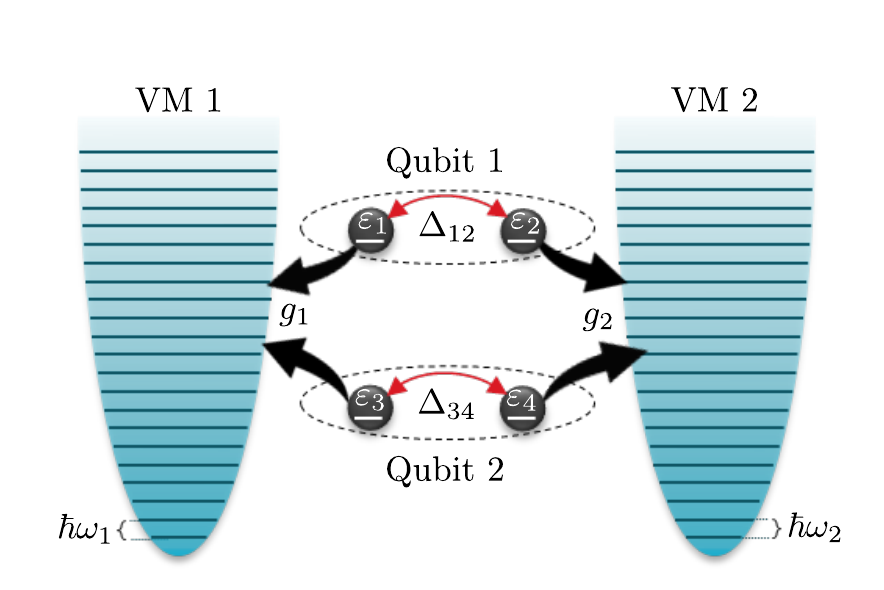}
\caption{Illustration of our system of interest: quantum dots $1$ and $3$ ($2$ and $4$)
are coupled with the vibrational (bosonic) mode VM1 (VM2) with frequency $\omega_{1(2)}$.
Additionally, the dots 1 (3) and 2 (4) are coupled by tunneling with real parameter $\Delta_{12(34)}$,
where $\varepsilon_i$ is the electronic level of the \emph{n}-th dot ($n=1,\ldots, 4$).
If two quantum dots share a single electron, the system can encode a qubit,
as shown by dashed lines.}
\label{fig:system}
\end{figure}

Our model consist of a multipartite system with two main parts, as illustrated in Fig.~\ref{fig:system}:
the electronic subspace $\mathcal{D}$ with two pairs
of quantum dots, each pair with potential to encode a qubit, and the subspace $\mathcal{V}$,
with two devices containing vibrational modes.
Here, the vibrational mode $1$ (VM1) couples to the electronic degrees of freedom of dots $1$ and $3$,
while the vibrational mode $2$ (VM2) couples to dots $2$ and $4$. Tunnelling is allowed between dots $1$ ($3$) and $2$ ($4$),
being responsible for flipping the electronic state, if a qubit is encoded on a pair of dots. That means that the electron-bosonic field interaction does not fulfill this role, in contrast to the Rabi model.

Concerning the computational basis, if we assume a single electronic level in a quantum dot, the elements of the basis have the general form $\ket{n_1 n_2 n_3 n_4}$, where each number indicates the occupation of the specific dot ($0$-empty and $1$-occupied). Additionally, the vibrational subspaces are spanned
by the occupation number basis states of the form $\ket{ml}_{\mathcal{V}}$, with $m,l=0,1,...,\infty$. Putting all together we end up with $ \ket{n_1 n_2 n_3 n_4}_{\mathcal{D}}\otimes\ket{ml}_{\mathcal{V}}=\ket{n_1 n_2 n_3 n_4, ml}$ as a general element of the computational basis used to span the complete space.

\subsection{General Hamiltonian}
\label{subsec:gen_model}
The Hamiltonian which describes the physical setup is written as
\begin{equation}
\label{eq:Hgeneral}
H=H_{\mathcal{D}}+H_{\mathcal{V}}+V_{\mathcal{DV}},
\end{equation}
where $H_{\mathcal{D}}$ and $H_{\mathcal{V}}$ are the free Hamiltonians of the quantum dots and vibrational modes subspaces,
respectively and $V_{\mathcal{DV}}$ is the dots-vibrational modes coupling.
The first term is written as:
\begin{eqnarray}
\label{eq:Hmol}
H_{\mathcal{D}}&=&\left[\sum_{i=1}^2\varepsilon_i N^{\mathcal{D}}_i+\Delta_{12}\left(S_1^\dagger S_2+S_2^\dagger S_1\right)\right]\\
&&+\left[\sum_{i=3}^4\varepsilon_i N^{\mathcal{D}}_i+\Delta_{34}\left(S_3^\dagger S_4+S_4^\dagger S_3\right)\right],\nonumber
\end{eqnarray}
where $S_{i}^\dagger$ ($S_{i}$) are the creation (annihilation) operators for the $i$-th
quantum dot and $N^{\mathcal{D}}_{i}=S_{i}^\dagger S_{i}$. The parameters $\varepsilon_{i}$ are the electronic levels for
each dot while $\Delta_{12(34)}$ is a real number describing the tunnel coupling.
If we consider a single vibrational mode per subsystem, the free Hamiltonian $ H_{\mathcal{V}}$ becomes ($\hbar=1$)
\begin{equation}
H_{\mathcal{V}} = \omega_{v_1} B_{v_1}^\dagger B_{v_1} + \omega_{v_2} B^\dagger_{v_2} B_{v_2},
\end{equation}
where $\omega_{v_j}$ is the energy of the corresponding $j$-th vibrational mode.
Here $B^{\dagger}_{v_j}$ ($B_{v_j}$) creates (annihilates) an excitation in a $j$-th vibrational mode subspace. Finally, the term $V_{\mathcal{DV}}$, which provides the electron-vibrational mode coupling, is written as
\begin{eqnarray}
\label{eq:Vmp}
V_{\mathcal{DV}}&=&g_1\left(N^{\mathcal{D}}_1+N^{\mathcal{D}}_3\right) \otimes \left({B}_{v_1}^\dagger+{B}_{v_1}\right)\nonumber\\
&&+g_2\left(N^{\mathcal{D}}_2+N^{\mathcal{D}}_4\right) \otimes  \left({B}_{v_2}^\dagger+{B}_{v_2}\right),
\end{eqnarray}
where parameter $g_v$ gives the coupling strength between electronic and vibrational degrees of freedom.

\subsection{Physical Parameters}
\label{subsec:parameters}
In our calculations, we will assume realistic parameters,
considering specifically the CNT scenario.
We set all physical quantities in terms of the energies $\omega_v$ of the vibrational modes,
once the comparisons of the tunneling and the electron-vibrational mode couplings in terms
of $\omega_v$ become the key ingredients behind the generation of the electronic entangled states.
We assume that both vibrational modes has the same frequency, i.e., $\omega_1=\omega_2=\omega$~\cite{walter2013},
and set the frequency at $\omega=20$ meV ($4.8$ THz), being in agreement with the high-frequencies values
of the radial breathing mode~\cite{leroy2004}. The tunneling parameter is fixed at $\Delta_{12}=\Delta_{34}=5\times 10^{-3}\omega=0.1$ meV,
as reported in experiments on parallel CNT quantum dots~\cite{Meyer13}.
In face of the potential of manipulation of the electron-phonon coupling in the context of quantum
dots and nanotubes~\cite{benyamini2014}, we are interested in tuning $g_v$
as done in some theoretical treatments~\cite{Sowa17,walter2013}, from 0.1$\omega$ up to 0.5$\omega$.
The temperature is assumed to be low enough to guarantee that the dominant vibrational state is at $\ket{0}\otimes\ket{0}$.
Typically, temperatures around $T=5$ K are used in experiments on transport in CNT~\cite{leroy2004}.
Finally, the values for the electronic levels, $\varepsilon_{i(j)}$ can be tunned via gate voltages
applied on the quantum dots arrangement~\cite{Shinkai09}.

\subsection{Effective Hamiltonians}
\label{subsec:effectiveH}
It is interesting to obtain further insight on the system by exploring some
analytical features of the full model. For instance, consider that the
electron-vibrational mode coupling is the same for both vibrational modes
so that, $g_1=g_2=g$~\footnote{As pointed out by Sowa \textit{et. al}~\cite{Sowa17},
there can be a phase difference in the coupling parameters $g_{v_j}$ given
by where $\phi_{v_j}=\mathbf{k}_{v_j} \cdot \mathbf{d}_{v_j}$, with $\mathbf{k}_{v_j}$
being the wavevector of the $j$ vibrational mode and $\mathbf{d}_{v_j}$
the distance between dots coupled with this specific mode.}.
In order to analyze the action of electron-vibrational mode and tunnel couplings,
we apply the Lang-Firsov~\cite{Mahanbook} unitary transformation over the
Hamiltonian in Eq.(\ref{eq:Hgeneral}). This transformation is regularly used in
the study of electron-phonon interaction, in the contexts of small-polaron
models~\cite{PhysRevLett.82.807} and strong correlated systems~\cite{kennes2017transient}.

We define the operator $S$ as
\begin{eqnarray}
S&=&\alpha \left(N^{\mathcal{D}}_1+N^{\mathcal{D}}_3\right) \otimes \left(B_{v_1}^\dagger - B_{v_1}\right)\nonumber\\
&&+\alpha \left(N^{\mathcal{D}}_2+N^{\mathcal{D}}_4\right)\otimes \left(B_{v_2}^\dagger - B_{v_2}\right),
\label{eq:Soperator}
\end{eqnarray}
with $\alpha=\frac{g}{\omega}$. The Lang-Firsov transformation consists on the calculation of the Hamiltonian $\bar{H} = e^{S} H e^{-S}$.
Considering $H$ as written in Eqs.(\ref{eq:Hgeneral}-\ref{eq:Vmp}), we find
\begin{equation}
\label{eq:Htrans}
\bar{H}=\left(\bar{H}_{\mathcal{D}}+V_{\mathrm{eff}}\right)+H_{\mathcal{V}}+\Delta^T_{\mathcal{DV}},
\end{equation}
where
\begin{equation}
\label{eq:Hmtrans}
\bar{H}_{\mathcal{D}}=\sum_{i=1}^4\widetilde{\varepsilon}_i N^{\mathcal{D}}_i,
\end{equation}
is the transformed Hamiltonian for the dots with $\widetilde{\varepsilon}_{i}$
being the shifted energy due to the action of the electron-vibrational mode coupling being
$\widetilde{\varepsilon}_{i}=\varepsilon_{i}-\alpha^2\omega$.
The term $V_{\mathrm{eff}}$ is an effective electron-electron interaction written as
\begin{equation}
\label{eq:Heeeff}
V_{\mathrm{eff}}=-2\alpha^2\omega N^{\mathcal{D}}_1\otimes N^{\mathcal{D}}_3-2\alpha^2\omega N^{\mathcal{D}}_2\otimes N^{\mathcal{D}}_4,
\end{equation}
and the last term
\begin{eqnarray}
\label{eq:Hvmvmeff}
&&\Delta^T_{\mathcal{DV}}=\left[\left(\Delta_{12} S_1^\dagger S_2\right) + \left(\Delta_{34}S_3^\dagger S_4\right)\right]\otimes \mathbb{D}_{12}\nonumber\\
&&\;\;+\left[\left(\Delta_{12} S_2^\dagger S_1 \right) + \left(\Delta_{34} S_4^\dagger S_3 \right)\right]\otimes \mathbb{D}^{\dagger}_{12},
\end{eqnarray}
describes an effective electron-vibrational mode coupling. Here we have define the operator
\begin{eqnarray}
\label{eq:dd12}
\mathbb{D}_{12}&=&e^{\alpha \left(B^\dagger_{v_1}-B_{v_1}\right)}\otimes e^{-\alpha \left(B^\dagger_{v_2}- B_{v_2}\right)}\nonumber\\
&=&D(\alpha)\otimes D(-\alpha),
\end{eqnarray}
which is a tensorial product of displacement operators, as defined for the quantum harmonic oscillator~\cite{Scullybook}.
The new transformed Hamiltonian, Eq.~(\ref{eq:Htrans}) and its terms Eqs.~(\ref{eq:Hmtrans})-(\ref{eq:Hvmvmeff}),
highlights important effects of the couplings considered on this particular physical system.
The first is a shift on the value of the electronic levels which depends on both,
the coupling parameter $g$ and $\omega$.
The second is the effective electron-electron interaction which couples the electrons from different qubits,
which is mediated by the electron-vibrational mode coupling.

Now, we are ready to encode two qubits in our physical system,
as sketched in Fig.~\ref{fig:system}.
We assume that each pair of quantum dots (dots 1-2 and dots 3-4)
contains a single electron. Therefore, we have a reduced electronic basis with four
states of the form $\ket{n_1 n_2 n_3 n_4}$, namely,
$\ket{1010}=\ket{\uparrow\uparrow}$, $\ket{1001}=\ket{\uparrow\downarrow}$,
$\ket{0110}=\ket{\downarrow\uparrow}$ and $\ket{0101}=\ket{\downarrow\downarrow}$,
where spin-$1/2$ notation was introduced.
In this restricted space, we use the matrix representations of operators $N^{\mathcal{D}}_{i}$
and $S_{i}$ to write the Hamiltonian, Eq.(\ref{eq:Hvmvmeff}), as
\begin{equation}\label{eq:qubitH}
 \bar{H}=\bar{H}_0+\bar{V},
\end{equation}
where
\begin{eqnarray}\label{eq:qubitH0}
\bar{H}_0&=&\left[\sum_{q=1}^{2}\frac{\delta_q}{2}\sigma_z^{(q)}-\alpha^2\omega\left(\sigma_z\otimes\sigma_z+I\right)\right]+H_{\mathcal{V}},
\end{eqnarray}
and
\begin{eqnarray}\label{eq:qubitV}
\bar{V}&=&\sum_{q=1}^{2}\Delta_{q}\left[\sigma_+^{(q)}\otimes \mathbb{D}_{12}+\sigma_-^{(q)}\otimes \mathbb{D}^{\dagger}_{12}\right],
\end{eqnarray}
where $q$ runs over the electronic qubits, so for qubit $q=1(2)$
the detuning is defined as $\delta_{1(2)}=\varepsilon_{1(3)}-\varepsilon_{2(4)}$
and the tunneling parameter is given by $\Delta_{1(2)}=\Delta_{12(34)}$. This particular form of our model is interesting, as it is able to reveal the emergence of an effective electronic interaction term, $\sigma_z\otimes\sigma_z$, in
a similar way to the models describing the experiments in charged QD~\cite{Shinkai09}.

\section{Spectral analysis}
\label{sec:eigenproblem}
We proceed to explore the characteristics of energy spectrum and
eigenstates of the Hamiltonian, Eq.(\ref{eq:Hgeneral}).
We focus on how the interplay between $g$, the tunnel coupling and the detuning $\delta_i$
can yield to the generation of maximally entangled states.
Along with the study of energy spectrum, we are interested on
the entanglement properties of the eigenstates.
It is well known that Coulomb interaction is behind the formation of
entangled states in coupled quantum dots molecule~\cite{Fujisawa11,Oliveira15}.
Because of the information provided by transformed Hamiltonian, Eq. (\ref{eq:Hmtrans}),
we expect the occurrence of signatures of the effective electron-electron
interaction on the entanglement degree of the eigenstates.

To quantify the entanglement degree, the measurement of concurrence is evaluated, as defined by Wootters~\cite{Wootters98},
which requires the calculation of the density matrix for each eigenstate.
We define $\hat{\rho}_l=\ket{\psi_l}\bra{\psi_l}$,
where $\ket{\psi_l}$ is the $l$-th eigenstate of Hamiltonian (\ref{eq:Hgeneral}).
Then, we calculate the reduced $4\times 4$ density matrix for the two qubits,
by tracing out the degrees of freedom of the vibrational modes
so $\hat{\rho}_{\mathcal{D},l}=\mathrm{Tr}_{\mathcal{V}}[\hat{\rho}_l]$.
An auxiliary Hermitian operator~\cite{Hill97}
is defined as $R_l=\sqrt{\sqrt{\hat{\rho}_{\mathcal{D},l}}\;\widetilde{\hat{\rho}_{\mathcal{D},l}}\sqrt{\hat{\rho}_{\mathcal{D},l}}}$,
where $\widetilde{\hat{\rho}_{\mathcal{D},l}}=(\sigma_y \otimes \sigma_y)\hat{\rho}^\star_{\mathcal{D},l}(\sigma_y \otimes \sigma_y)$,
is the spin-flipped matrix with $\hat{\rho}^\star_{\mathcal{D},l}$
being the complex conjugate of $\hat{\rho}_{\mathcal{D},l}$.
Finally, the concurrence is calculated considering $C=\mathrm{max}(0,\lambda_1-\lambda_2-\lambda_3-\lambda_4)$
where $\lambda_k$ ($k=1,\ldots, 4$) is the \emph{k}-th eigenvalue of the
operator $R_l$ in decreasing order.

Motivated by our previous work~\cite{Oliveira15}, which demonstrated that, by changing $\delta_q$, it is possible to reach maximally entangled states in Coulomb interacting qubits, here we check the behavior of energy and concurrence as a function of $\delta_1$, for different
values of $\delta_2$. To begin our analysis, it is instructive to write the free energies of the two qubit model, Eq.~(\ref{eq:qubitH0}),
neglecting the tunnel coupling ($\Delta_q=0$), which results in the following expressions:
\begin{eqnarray}
 \varepsilon_{\uparrow\uparrow,ml}&=&\frac{\delta_1+\delta_2}{2}-2\alpha^2\omega+(m+l)\omega\nonumber\\
 \varepsilon_{\downarrow\downarrow,ml}&=&-\left(\frac{\delta_1+\delta_2}{2}\right)-2\alpha^2\omega+(m+l)\omega\nonumber\\
 \varepsilon_{\uparrow\downarrow,ml}&=&\frac{\delta_1-\delta_2}{2}+(m+l)\omega\nonumber\\
 \varepsilon_{\downarrow\uparrow,ml}&=&-\left(\frac{\delta_1-\delta_2}{2}\right)+(m+l)\omega.
 \label{eq:freeenergies}
\end{eqnarray}
The above equations allow the discussion of some important
features concerning the spectrum, which shed light on the
conditions for the formation of maximally entangled states.
Notice that $\varepsilon_{\uparrow\uparrow,ml}=\varepsilon_{\downarrow\uparrow,ml}$
at $\delta_1=2\alpha^2\omega$ and $\varepsilon_{\downarrow\downarrow,ml}=\varepsilon_{\uparrow\downarrow,ml}$
at $\delta_1=-2\alpha^2\omega$. At these $\delta_1$ values we expect the appearance of
anticrossings due to the tunneling term of the Hamiltonian,
that couples states like $\ket{\uparrow\uparrow,00}$ to $\ket{\downarrow\uparrow,00}$
and $\ket{\downarrow\downarrow,00}$ to $\ket{\uparrow\downarrow,00}$.
However, in terms of entanglement, it is expected a low value of concurrence,
once the eigenstates would be separable. Additionally, we find $\varepsilon_{\uparrow\downarrow,ml}=\varepsilon_{\downarrow\uparrow,ml}$
at $\delta_1=\delta_2$ and $\varepsilon_{\uparrow\uparrow,ml}=\varepsilon_{\downarrow\downarrow,ml}$
at $\delta_1=-\delta_2$. The Hamiltonian does not provide a first order coupling
between states such as $\ket{\uparrow\downarrow,00}$ and $\ket{\downarrow\uparrow,00}$.
However, these states can be coupled via second order transitions.
This means that if the system is initialized at $\ket{\uparrow\downarrow,00}$,
it can evolve to $\ket{\downarrow\uparrow,00}$, passing through
intermediate states such as $\ket{\uparrow\uparrow,ml}$ and
$\ket{\downarrow\downarrow,ml}$. At this specific condition, highly entangled
states can be formed due to virtual processes. This effect will be used
to dynamically generate entangled states, as discussed in the next section.

\begin{figure}[ht]
\centering\includegraphics[width=1\linewidth]{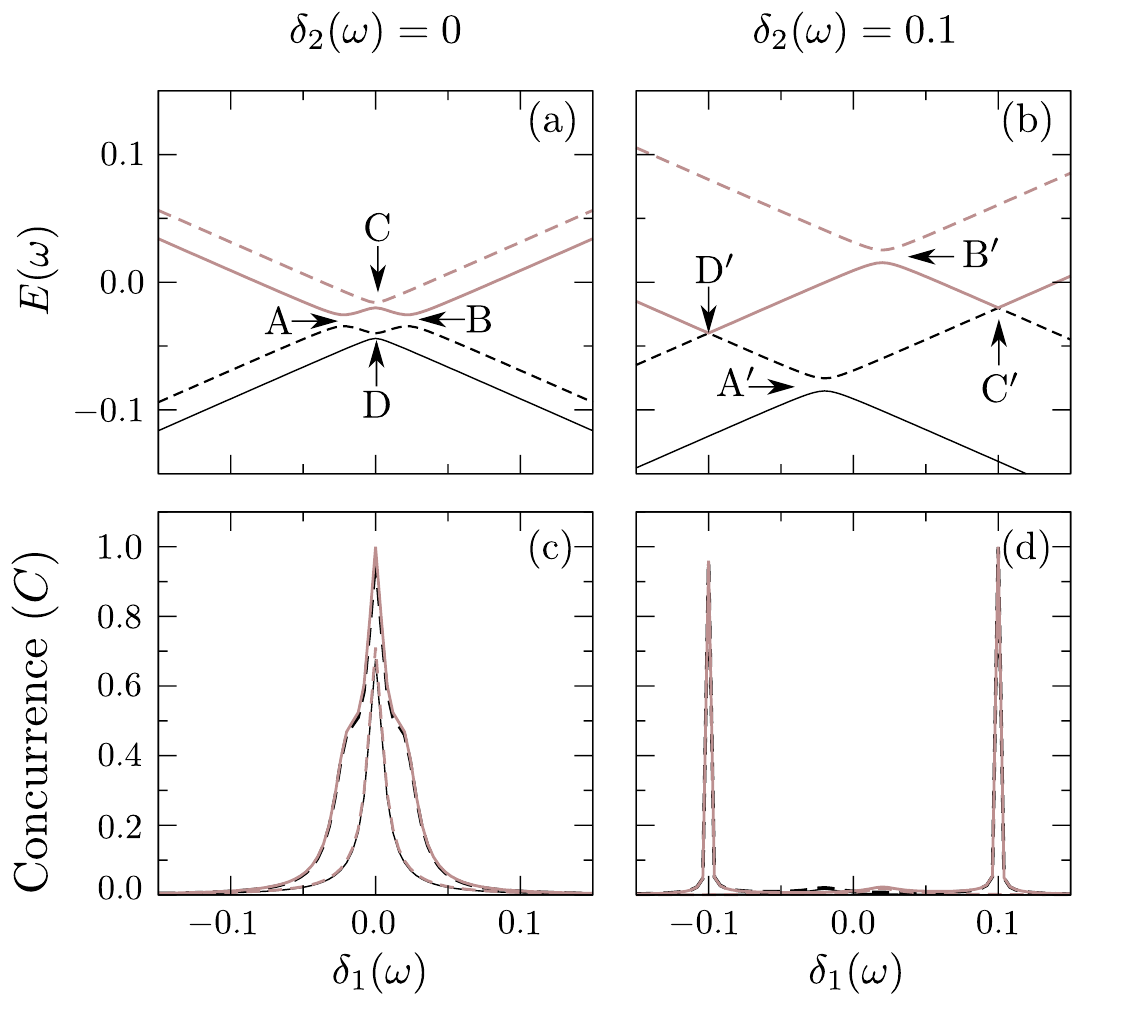}
\caption{Spectrum and entanglement degree as functions of $\delta_1$, of the first (solid black line),
the second (dashed black line), the third (solid brown line) and the four (dashed brown line)
eigenstate of the general Hamiltonian, Eq.(\ref{eq:Hgeneral}). Here, it is shown (a) the energies
and (b) the concurrence, considering $\delta_2=0$ and (c) energies and (d) concurrence,
for $\delta_2=0.1\omega$. Anticrossings are indicated by arrows and identified with letters
(see main text for details). Physical parameters are $g=0.1\omega$ and $\Delta_{1}=\Delta_{2}=5\times 10^{-3}\omega$. }
\label{fig:eigenproblem}
\end{figure}

Fig.~\ref{fig:eigenproblem} shows the first four eigenvalues of the general Hamiltonian, Eq.(\ref{eq:Hgeneral}),
and the corresponding concurrences as functions of $\delta_1$,
considering two different values of $\delta_2$. To perform our numerical calculation,
both bases associated with the vibrational modes
are truncated at $m_{\mathrm{max}}=l_{\mathrm{max}}=13$. This number of computational states
is enough to guarantee the accuracy of the lower eigenenergies.
With respect to the energies, the results reported on Fig.~\ref{fig:eigenproblem}(a)-(b) show
anticrossings labeled with letters (A,B) and
(A$^{\prime}$,B$^{\prime}$), corresponding to first order transitions
that play a role at $\delta_1=\pm 2\alpha^2\omega$. In contrast, the smaller anticrossings indicated by letters (C,D)
and (C$^{\prime}$,D$^{\prime}$) are associated with second and higher order transitions that take place at $\delta_1=\pm \delta_2$.

\begin{figure}[ht]
\centering\includegraphics[width=1\linewidth]{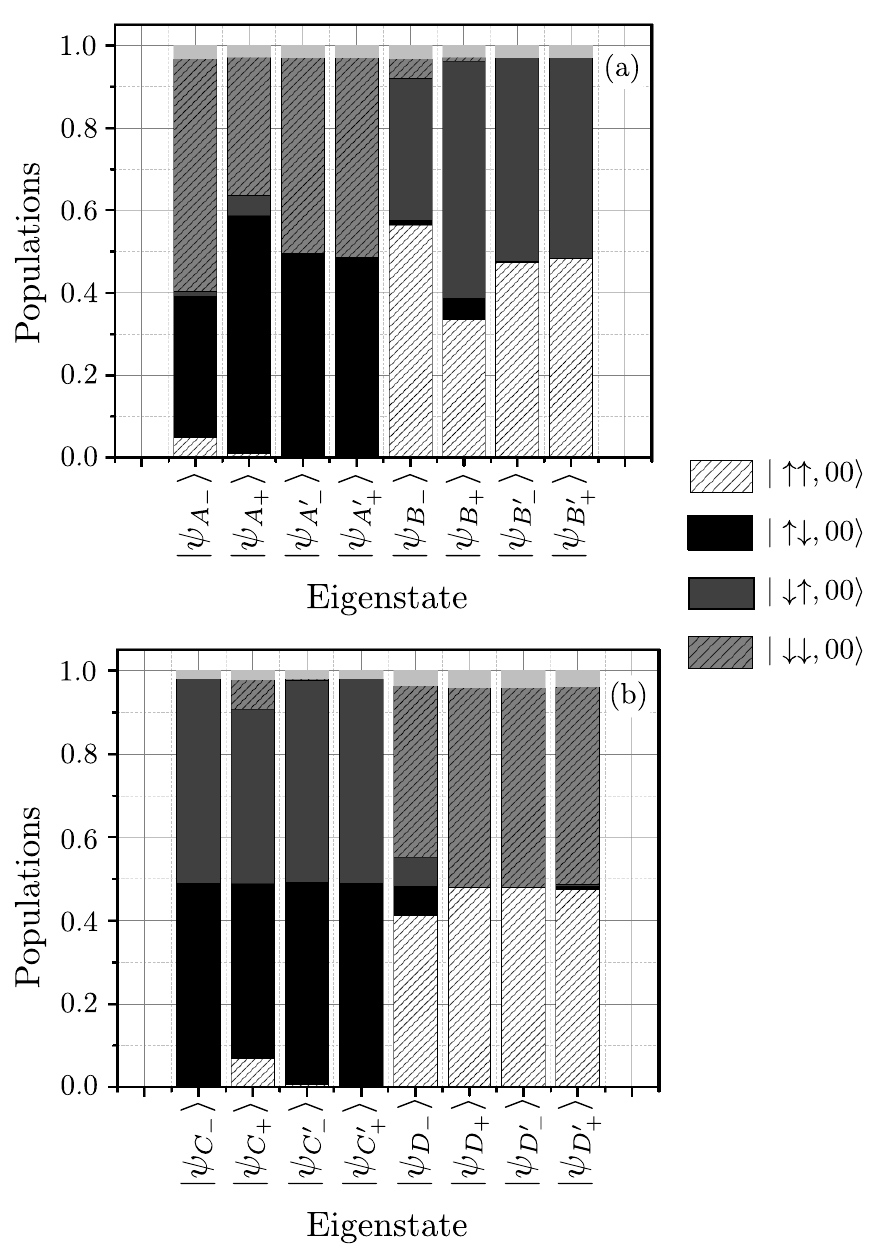}
\caption{Stacked bar graph showing the populations of
states of the computational basis for the eigenstates at the anticrossings positions at
Fig.~\ref{fig:eigenproblem} of: (a) first order and (b) second order .
Color and patterns for the main contributors (states with $m=l=0$) are described on the right side of the panel.
Solid gray is used for the sum of the populations of other states.
Physical parameters are $g=0.1\omega$ and $\Delta_{1}=\Delta_{2}=5\times 10^{-3}\omega$.}
\label{fig:eigenstates}
\end{figure}

To understand what happens with the eigenvectors corresponding to the first order (A,B)-(A$^{\prime}$,B$^{\prime}$) and the second order anticrossings
(C,D)-(C$^{\prime}$,D$^{\prime}$), we use an stacked bar graph, Fig.~\ref{fig:eigenstates}(a)-(b).
Each color and patterns corresponds to the population of a respective state of the 4D basis given by $\left\{\ket{\uparrow\uparrow,00},\ket{\uparrow\downarrow,00},\ket{\downarrow\uparrow,00},\ket{\downarrow\downarrow,00}\right\}$, as indicated in the figure. The gray bar gives the sum of the populations of the remain components with at least one vibrational mode excitation.
Comparing both types of anticrossings, we verified that the eigenstates in Fig.~\ref{fig:eigenstates}(a)
are mainly superpositions of states, with at least one spin component at the same orientation,
such as $\ket{\uparrow\downarrow,00}$ and $\ket{\downarrow\downarrow,00}$,
or $\ket{\downarrow\uparrow,00}$ and $\ket{\uparrow\uparrow,00}$,
which results in low entanglement. For clarity, in Appendix~\ref{ap2:anticrossings} we show the expansions of
the eigenstates in the computational basis. In contrast, in Fig.~\ref{fig:eigenstates}(b)
we find eigenstates such as $\ket{\psi_{C-}}$, $\ket{\psi_{D+}}$, $\ket{\psi_{C^{\prime}\pm}}$ and $\ket{\psi_{D^{\prime}\pm}}$
that are highly entangled eigenstates. For instance, $\ket{\psi_{C^{\prime}+}}$ can be written with
good accuracy as
\begin{equation}
 \ket{\psi_{C^{\prime}+}}=\frac{1}{\sqrt{2}} \left(\ket{\uparrow\downarrow}+e^{i\varphi} \ket{\downarrow\uparrow}\right) \otimes \ket{00},
\end{equation}
where $\varphi$ is a relative phase.
This shows that the electron-vibrational mode coupling
is the source of the emergence of electronic Bell states as eigenstates.

In Fig.~\ref{fig:eigenproblem}(c)-(d), we show the behavior of concurrence as a function
of $\delta_1$. Note that $C$ reaches values close to one,
corresponding to the anticrossings (C, D) and (C$^{\prime}$, D$^{\prime}$).
This is consistent with the eigenstates $\ket{\psi_{C-}}$, $\ket{\psi_{D+}}$
for $\delta_2=0$, and $\ket{\psi_{C^{\prime}\pm}}$ and $\ket{\psi_{D^{\prime}\pm}}$ for $\delta_2=0.1\omega$,
shown in Fig.~\ref{fig:eigenstates}.
The eigenstate $\ket{\psi_{C^{\prime}+}}$, in particular, presents the larger value
of concurrence, $C \approx 1$.
This fact is in agreement with an analytical solution (Appendix~\ref{ap1:bellboson})
of the matricial representation of the general Hamiltonian, Eq.(\ref{eq:Hgeneral}), in a rotated electronic basis of Bell states.
Also, it is worthy to note that the condition $\delta_2=0.1\omega$,
which results on the anticrossing C$^{\prime}$, favors an energetically isolated two-level
subspace within $\left\{\ket{\uparrow\downarrow,00},\ket{\downarrow\uparrow,00}\right\}$,
which will be used in the next section in order to find an effective two-level model.
Note also that we find $C\approx 0.9$ at anticrossings D and D$^{\prime}$.
By checking the values of the eigenstates coefficients, shown in Appendix~\ref{ap2:anticrossings},
we verified that although the electronic part is roughly similar to the Bell
states $\ket{\Phi_{\pm}}$, as defined in Eq.(\ref{eq:ap1_bellstates}), the superposition has contributions from other electronic states, thus suppressing the degree of entanglement.

\section{Dynamical generation of maximally entangled electronic states}
\label{sec:dynamics}

After studying the properties of the eigenstates of the model,
we are ready to explore the generation of electronic entangled states
by quantum dynamics with the general form,
\begin{subequations}
\begin{align}
\ket{\Psi(\varphi)}=\frac{1}{\sqrt{2}} \left(\ket{\uparrow\downarrow}+e^{i\varphi} \ket{\downarrow\uparrow}\right),
\label{eq:Bellpsi}\\
\ket{\Phi(\vartheta)}=\frac{1}{\sqrt{2}} \left(\ket{\uparrow\uparrow}+e^{i\vartheta} \ket{\downarrow\downarrow}\right),
\label{eq:Bellphi}
\end{align}
\end{subequations}
where $\varphi$ and $\vartheta$ are relative
phases \footnote{If we consider the values $\varphi,\vartheta=\{0,\pi\}$
we obtain the orthonormal basis for the $4$D space given by the Bell
states $\ket{\Psi_{\pm}}$ and $\ket{\Phi_{\pm}}$, defined in Eq.{\ref{eq:ap1_bellstates}}}.

First, we obtain numerically the density operator $\rho(t)=\ket{\psi(t)}\bra{\psi(t)}$,
using the Hamiltonian (\ref{eq:Hgeneral}), considering a specific initial state.
Then, by tracing out the vibrational degrees of freedom,
we obtain the electronic reduced density matrix
\begin{equation}
 \rho_{\mathcal{D}}(t)=\mathrm{Tr}_{\mathcal{V}}[\rho(t)].
 \label{eq:rhodots}
\end{equation}
The equation above is used to explore the dynamical behavior of the electronic part of the system,
through the analysis of the evolution of the concurrence and the fidelity of the evolved state.
The system is initialized at $\rho_0=\ket{\uparrow\downarrow,00}\bra{\uparrow\downarrow,00}$,
where the vibrational part is experimentally feasible at low temperature (Sec.~\ref{subsec:parameters}).
This choice of $\rho_0$ is motivated by our findings concerning
the anticrossing C$^{\prime}$, which favors the generation
of maximally entangled electronic states $\ket{\Psi(\varphi)}$.
Analogously, the choice of $\ket{\uparrow\uparrow,00}$ or $\ket{\downarrow\downarrow,00}$
as initial states could result on the formation of entangled states of the form $\ket{\Phi(\vartheta)}$,
for a dynamics considering the specific conditions of anticrossing D$^{\prime}$.
Any choice of initialization for the electronic state is realistic, once the experimental
setup can be coupled to a set of auxiliary sources and drains of electrons,
allowing charge injection at any of the quantum dots on the physical system.
\begin{figure}[hb]
\centering\includegraphics[width=1\linewidth]{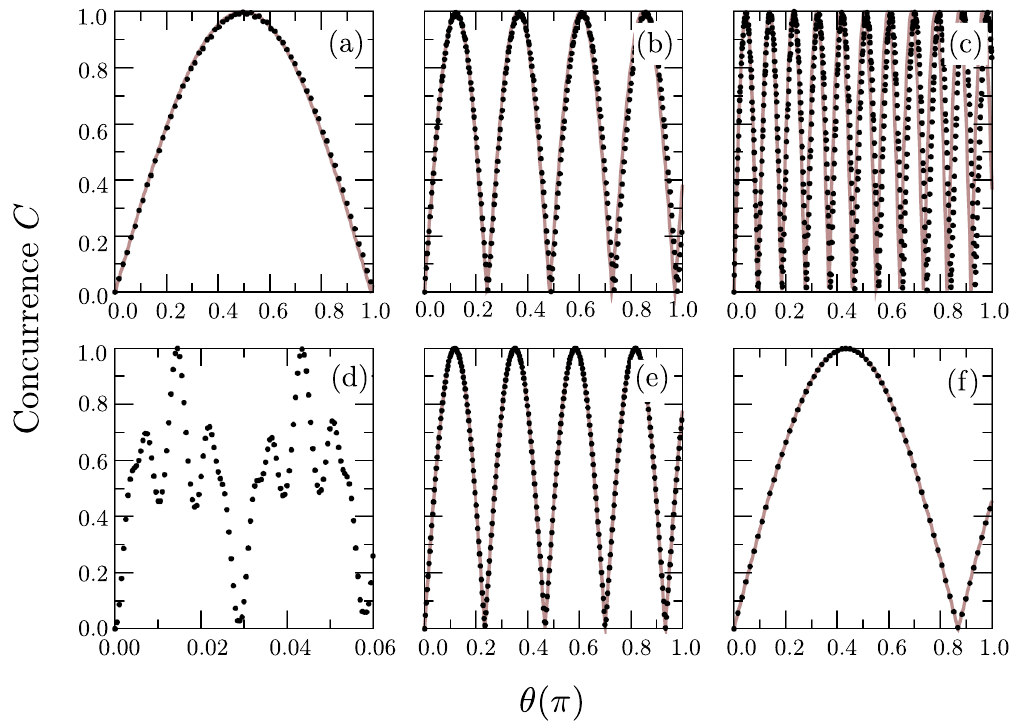}
\caption{Dynamics of the entanglement degree of the electrons as function of $\theta$
for different values of $g$, considering the physical parameters of the anticrossing C$^{\prime}$:
$\Delta=\Delta_{1}=\Delta_{2}=5\times 10^{-3}\omega$ and $\delta=\delta_1=\delta_2=0.1\omega$.
Here, the numerical calculation of $C$ (black dots) and the analytic expression $C_{\mathrm{2ls}}(t)$
as defined in Eq.~(\ref{eq:Cexact}) (brown line) are plotted considering (a) $g=0.05\omega$,
(b) $g=0.10\omega$, (c)$g=0.15\omega$, (d) $g=0.22\omega$, (e) $g=0.40\omega$, and (f) $g=0.50\omega$.}
\label{fig:dynconcurrence}
\end{figure}

Black dots in Fig.~\ref{fig:dynconcurrence} show our numerical results for the entanglement dynamics
considering the specific choice of parameters associated with the anticrossing C$^{\prime}$,
being $\Delta=\Delta_{1}=\Delta_{2}$ and $\delta_1=\delta_2=0.1\omega$,
and different values of the electron-vibrational mode coupling $g$. We define a new time-dependent variable,
$\theta=\Omega_0 t$, with $\Omega_0=2|\Delta|^2/\omega$, which will prove to be very useful in the discussion of the quantum dynamics of our system, as it will be noticed in the following discussions. Comparing the six panels, we notice that the concurrence $C$ shows an oscillatory behavior, with its maximum value reaching $C=1$, indicating the formation of maximally entangled electronic states.
Additionally, the period of the oscillations has a non-monotonic behavior as $g$ increases.
We identify three different situations: (i) for a first range of values of $g$ ($g<0.2\omega$)
the system performs sinusoidal oscillations with a decreasing period as $g$ increases,
as can be seen in Figs.~\ref{fig:dynconcurrence}(a)-(c); (ii) there is an intermediate
range of values of $g$ where the dynamics do not correspond to a sinusoidal function,
although they oscillate in a time scale significatively shorter (by two orders of magnitude)
than the other cases, as illustrated by Fig.~\ref{fig:dynconcurrence}(d) with $g=0.22\omega$;
(iii) after this intermediate regime, the system performs sinusoidal oscillations again,
although its period increases as $g$ increases as can be verified from  Figs.~\ref{fig:dynconcurrence}(e)-(f).

Our goal is to understand the physical description associated to the behavior of the concurrence described above.
From our discussion in Sec.~\ref{sec:eigenproblem},
in view of the characteristics of the anticrossing C$^{\prime}$,
we expect a dynamics associated with a two-level subspace with the elements being
the states $\ket{\mathrm{I}}=\ket{\uparrow\downarrow,00}$ and $\ket{\mathrm{II}}=\ket{\downarrow\uparrow,00}$.
Because the effective coupling between $\ket{\mathrm{I}}$ and $\ket{\mathrm{II}}$
involves second order transitions, we estimate its value applying the perturbation theory,
by calculating the matrix element $\Omega=\bra{\mathrm{I}}H_{\mathrm{eff}}\ket{\mathrm{II}}$ defined as
\begin{equation}
\Omega=\bra{\mathrm{I}}\bar{H}_0\ket{\mathrm{II}}+\sum_{\sigma=\uparrow}^{\downarrow}\sum^{\infty}_{m,l=0}\frac{\mathcal{V}_{\mathrm{I},\mathrm{II}}(\sigma,ml)}
{\varepsilon_{\downarrow\uparrow,00}-\varepsilon_{\sigma\sigma,ml}},
\label{eq:Heff2L}
\end{equation}
where
\[\mathcal{V}_{\mathrm{I},\mathrm{II}}(\sigma,ml)=\bra{\mathrm{I}}\bar{V}\ket{\sigma\sigma,ml}\bra{\sigma\sigma,ml}\bar{V}\ket{\mathrm{II}},\]
We expect that $\Omega$ provides the characteristic frequencies found in Fig.~\ref{fig:dynconcurrence},
which is valid for small values of the coupling $g$ ($g\ll\omega$).
The calculation requires the use of a well known property of the displacement operator~\cite{Scullybook} given by
\begin{eqnarray}
\bra{ml}\mathbb{D}_{12}\ket{00}&=&\bra{m}D_1(\alpha)\ket{0}\bra{l}D_2(-\alpha)\ket{0}\nonumber\\
&=&e^{-\alpha^2}\frac{\alpha^m}{\sqrt{m!}}\frac{(-\alpha)^l}{\sqrt{l!}}.
\end{eqnarray}
After a straightforward calculation, we arrive in the following expression for $\Omega$
\begin{equation}
\Omega=\Omega_0 e^{-2\alpha^2} \sum_{m,l=0}^\infty \frac{\alpha^{2m}}{m!} \frac{\alpha^{2l}}{l!} \frac{[2\alpha^2-(m+l)]}{[2\alpha^2-(m+l)]^2-(\frac{{\delta}}{\omega})^2},
\label{eq:bellfrequency}
\end{equation}
where $\Omega_0=2|\Delta|^2/\omega$. This effective coupling parameter describes a second order tunneling process, mediated by the electron-vibrational mode interaction.

\begin{figure}[tb]
\centering\includegraphics[width=0.8\linewidth]{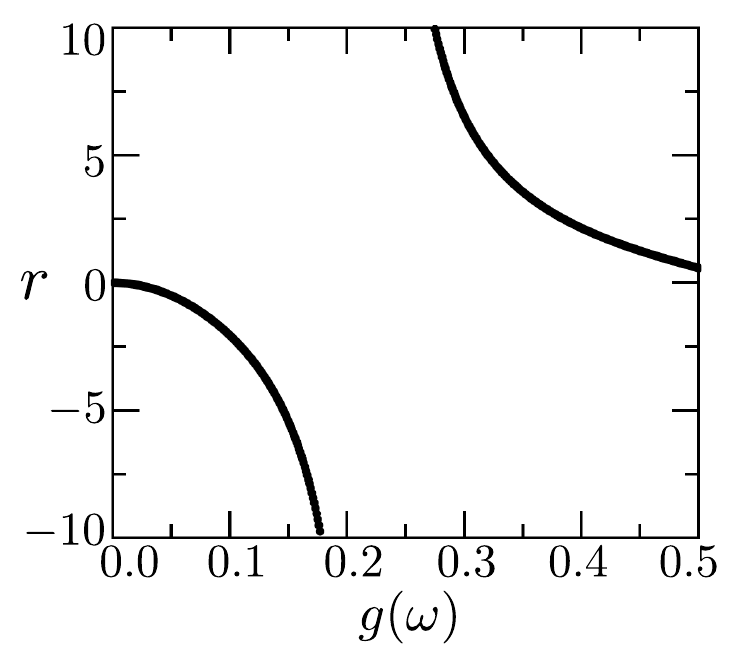}
\caption{Ratio $r=\Omega/\Omega_0$ as a function of the electron-vibrational mode coupling, $g$,
as given by Eq. (\ref{eq:bellfrequency}), considering the physical parameters
of the anticrossing C$^{\prime}$: $\Delta=\Delta_{1}=\Delta_{2}=5\times 10^{-3}\omega$ and $\delta=\delta_1=\delta_2=0.1\omega$.}
\label{fig:frequencyratio}
\end{figure}

To check the behavior of $\Omega$, we plot in Fig.~\ref{fig:frequencyratio} the ratio $r=\Omega/\Omega_0$ as function of coupling $g$, considering the physical conditions associated with anticrossing C$^{\prime}$.
From our results, we are able to identify three different behaviors: (i) at small values of $g$, between $g\approx 0$ and $g < 0.18 \omega$, $|r|$ increases as $g$ increases; (ii) if $g\approx0.22\omega$, the factor $2\alpha^2$ becomes comparable with $\delta/\omega$ so the denominator on Eq. (\ref{eq:Heff2L}) goes to infinity if $m=l=0$.
By checking the exact dynamics at this values of $g$, we realize that the state $\varepsilon_{\uparrow\uparrow,00}$ becomes resonant with $\varepsilon_{\uparrow\downarrow,00}$
and $\varepsilon_{\downarrow\uparrow,00}$ so the system evolves to a superposition of this three states and the effective two-level model is no longer valid;
(iii) for $0.2\omega<g<0.5\omega$, the value of $|r|$ decreases as $g$ increases.

At this point, we search for a more detailed characterization of the sinusoidal oscillations.
Calculating the evolved state, considering an effective two-level system ($\mathrm{2ls}$)
described by $H_{\mathrm{2ls}}=\Omega\ket{\mathrm{I}}\bra{\mathrm{II}}+\mathrm{h.c.}$,
and the same initial condition used in the numerical analysis, $\ket{\psi_0}=\ket{\uparrow\downarrow,00}=\ket{I}$,
we obtain
\begin{equation}
\ket{\psi(t)}_{\mathrm{2ls}}=\cos{\Omega t}\ket{\mathrm{I}}+e^{-i\pi/2}\sin{\Omega t}\ket{\mathrm{II}},
\label{eq:psit_2ls0}
\end{equation}
which, in terms of $r$ and $\theta$, is written as
\begin{equation}
\ket{\psi(t)}_{\mathrm{2ls}}=\left[\cos{(r\theta)}\ket{\uparrow\downarrow}+e^{-i\pi/2}\sin{(r\theta)}\ket{\downarrow\uparrow}\right]\otimes\ket{00}.
\label{eq:psit_2ls}
\end{equation}
The analytical expression of the concurrence, considering the effective two-level model,
takes the form~\cite{Hill97,PhysRevA.54.3824}
\begin{equation}
C_{\mathrm{2ls}}(r)=2|\cos(r \theta) \sin(r \theta)|.
\label{eq:Cexact}
\end{equation}

In Fig.~(\ref{fig:dynconcurrence}),
the behavior of the $C_{\mathrm{2ls}}$ is shown using solid gray lines. Notice that Eq.(\ref{eq:Cexact}) is in good agreement with the full numerical calculations of concurrence.
That means that the effective two-level model is able to describe the non-monotonic behavior of the sinusoidal oscillations of the concurrence. In the intermediate range of $g$'s values, this simplified model does not apply, as $\left|\Omega\right|\rightarrow\infty$. Interestingly, though, it catches the fast oscillations observed in Fig.~\ref{fig:dynconcurrence}(d).

To complete our discussion, we compute the fidelity of the electronic state given by
\begin{equation}
\mathcal{F}(t)=\mathrm{Tr}_{\mathcal{D}}[\rho_{\mathcal{D}}(t)\rho_{\mathcal{D}}^{\mathrm{tar}}],
\label{eq:fidelity}
\end{equation}
where $\rho_{\mathcal{D}}^{\mathrm{tar}}=\ket{\Psi(\varphi_{\mathrm{tar}})}\bra{\Psi(\varphi_{\mathrm{tar}})}$, where $\ket{\Psi(\varphi_{\mathrm{tar}})}$ is defined in Eq.(\ref{eq:Bellpsi}), being $\varphi_{\mathrm{tar}}$ the relative phase of a specific target state .
In our simulations, we choose $g=0.1\omega$ and $g=0.4\omega$,
to explore one example of each range of $g$ with sinusoidal oscillations of Fig.~\ref{fig:dynconcurrence}.

\begin{figure}[tb]
\centering\includegraphics[width=1\linewidth]{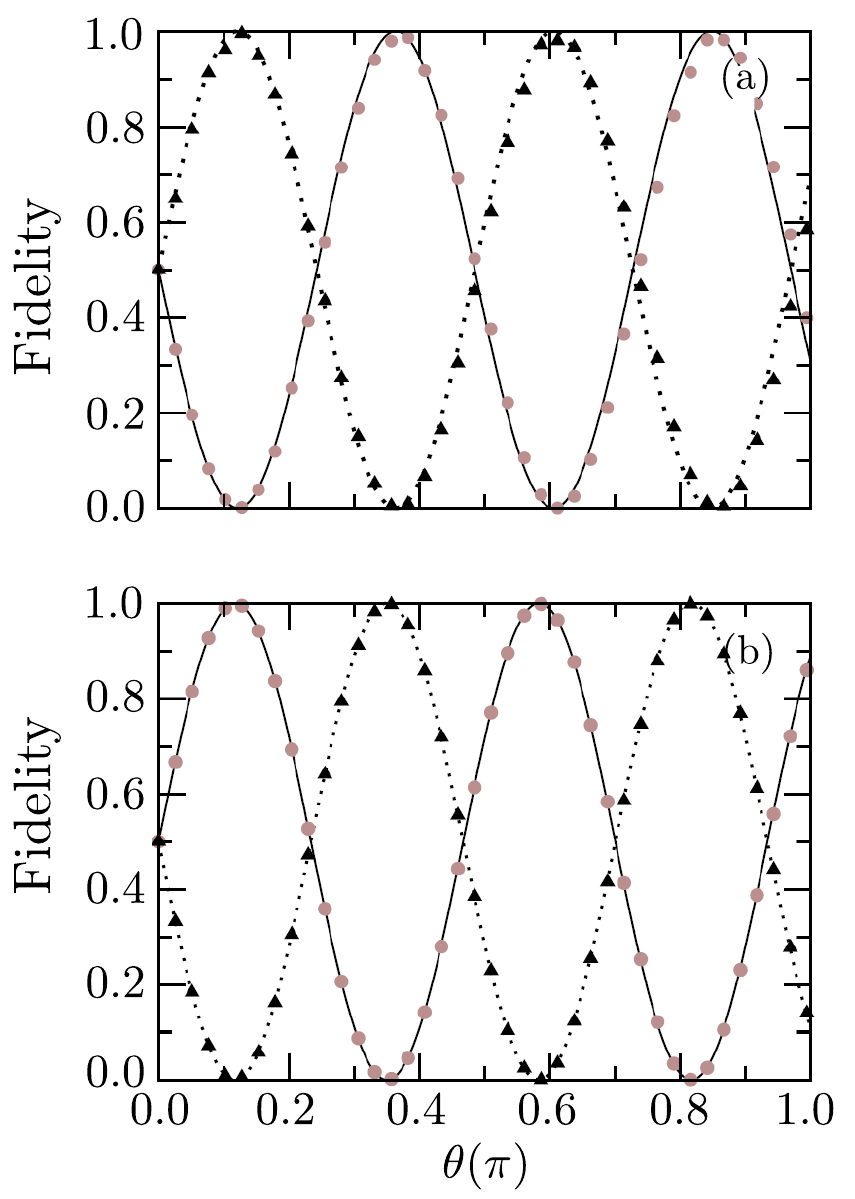}
\caption{Evolution of the fidelity, $\mathcal{F}$, considering the target state $\ket{\Psi(\varphi_{\mathrm{tar}})}$ for two choices of relative phase, $\varphi_{\mathrm{tar}}=-\pi/2$ (brown dots), and $\varphi_{\mathrm{tar}}=\pi/2$ (black triangles). The physical parameters correspond to those considered in Fig.~\ref{fig:dynconcurrence} with (a) $g=0.1\omega$ and (b) $g=0.4\omega$. Lines show the evolution of $ \mathcal{F}_{\mathrm{2ls}}(\theta)$ for the same physical conditions considering $\varphi_{\mathrm{tar}}=-\pi/2$ (solid black line) and $\varphi_{\mathrm{tar}}=\pi/2$ (dotted black line).}
\label{fig:fidelity}
\end{figure}

Our results are shown in Fig.~\ref{fig:fidelity}, considering two different values for relative phase
of the target state: $\varphi_{\mathrm{tar}}=-\pi/2$ (brown dots)
and $\varphi_{\mathrm{tar}}=\pi/2$ (black triangles).
Notice that the fidelity for both cases
of $\varphi_{\mathrm{tar}}$ oscillates out of phase between $0$ and $1$, and the comparison
between them permits to describe accurately the electronic dynamics. Specifically, in Fig.~\ref{fig:fidelity}(a)
considering $g=0.1\omega$,  the initial state $\ket{\uparrow\downarrow}$
evolves to a maximally entangled state of the form $\ket{\Psi(\varphi)}$,
which alternates between $\ket{\Psi(\pi/2)}$, at $r\theta=\pi/4$, and $\ket{\Psi(-\pi/2)}$,
at $r\theta=3\pi/4$. Analogously, the results for $g=0.4\omega$,
Fig.~\ref{fig:fidelity}(b), exhibit the same oscillations,
although they are out of phase if compared with Fig.~\ref{fig:fidelity}(a).

The differences between Fig.~\ref{fig:fidelity}(a) and Fig.~\ref{fig:fidelity}(b)
are explained by the behavior of ratio $r$, which goes from negative to positive
value depending on the value of $g$.
Considering the evolved state associated with the effective two-level model,
Eq.(\ref{eq:psit_2ls}), a change on the sign of $r$ implies
in a change of the relative phase of the evolved state.
Calculating the fidelity of the analytical solution given by Eq.(\ref{eq:psit_2ls}),
considering the same target state $\ket{\Psi(\varphi_{\mathrm{tar}})}$,
it reads as
\begin{equation}
 \mathcal{F}_{\mathrm{2ls}}(r\theta)=\frac{1}{2}-\cos(r\theta)\sin(r\theta)\sin(\varphi_{\mathrm{tar}}).
\end{equation}
The evolution of this function is illustrated with the lines in Fig.~\ref{fig:fidelity}, for each case of $g$ and $\varphi_{\mathrm{tar}}$, showing good agreement with the numerical results.

\section{Feasibility of the generation of electronic entangled states}
\label{sec:feasibilty}

In this section, we review some aspects about physical parameters for the generation of electronic maximally entangled states. The Eq.(\ref{eq:Cexact}) shows that the concurrence $C$ reaches its first maximum value at $r \theta_{\mathrm{max}} = \pi/4$,
which in terms of time scale results in $t_{\mathrm{max}}\approx 1$ ns, for $g=0.05\omega$ in
Fig.~\ref{fig:dynconcurrence}(a), reducing to $t_{\mathrm{max}}\approx 0.1$ ns,
for $g=0.15\omega$ in Fig.~\ref{fig:dynconcurrence}(c). To check the robustness of the generation of the entangled states, here we
consider the effect of charge dephasing, which is the main mechanism of decoherence
in our physical system~\cite{Zhang18}, once our proposal requires low
temperatures and small values of excitations on the vibrational modes.
We simulate this process solving a master equation~\cite{Fujisawa11},
where Lindblad operators are introduced to take into account
the dephasing of the coherent oscillations in each qubit.
The time scale of the dephasing process
is given by $T_{\mathrm{deph}}=1/\gamma_{\mathrm{deph}}=h/\Gamma_{\mathrm{deph}}$,
where $\Gamma_{\mathrm{deph}}$ is the dephasing rate,
in energy units.

\begin{figure}[tb]
\centering\includegraphics[width=1.05\linewidth]{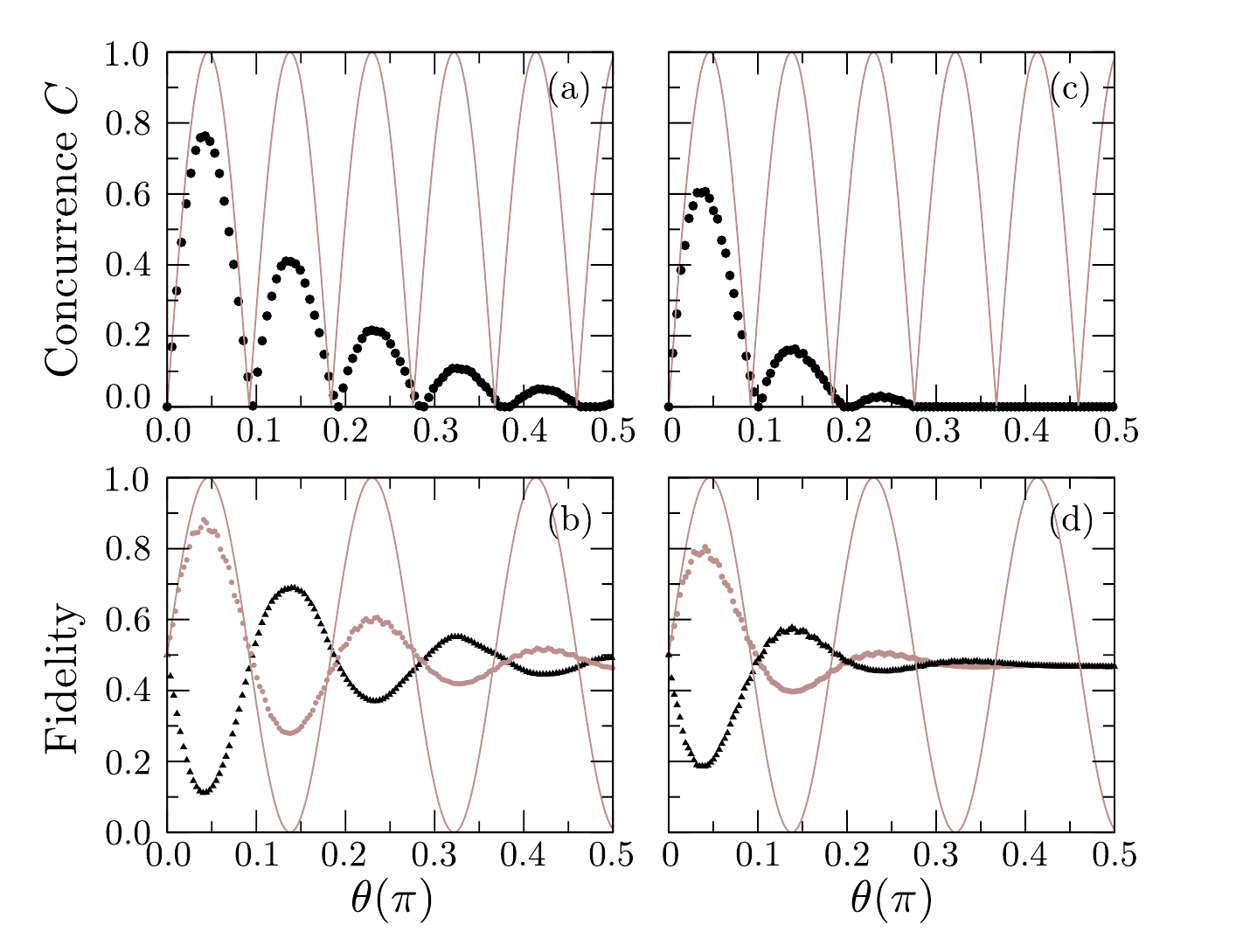}
\caption{Evolution of the concurrence and the fidelity, assuming two different values of charge dephasing rate $\Gamma_{\mathrm{deph}}$, calculated by solving a Lindblad master equation, with $\omega=20$ meV, $\Delta=5\times 10^{-3}\omega=0.1$ meV and $\delta_1=\delta_2=0.1\omega=2$ meV, with $g=0.15\omega=3$ meV corresponding to the case illustrated in panel (c) in Fig~\ref{fig:dynconcurrence}. (a) Concurrence $C$ (black dots) for $\Gamma_{\mathrm{deph}}=1\times 10^{-4}\omega=20\times 10^{-2}$ meV; (b) fidelity considering the target state $\ket{\Psi(\varphi_{\mathrm{tar}})}$ for two choices of relative phase, $\varphi_{\mathrm{tar}}=-\pi/2$ (brown dots), and $\varphi_{\mathrm{tar}}=\pi/2$ (black triangles), considering the same value $\Gamma_{\mathrm{deph}}$ in panel (a); (c)  Concurrence $C$ (black dots) for $\Gamma_{\mathrm{deph}}=2\times 10^{-4}\omega=40\times 10^{-2}$ meV; (d) Fidelity, also for $\Gamma_{\mathrm{deph}}=2\times 10^{-4}\omega$ using the same scheme of colors and symbols of panel (b). In all panels, we illustrate the behavior when the dephasing is neglected (brown solid line).}
\label{fig:dephasing}
\end{figure}

In Fig.~\ref{fig:dephasing} we show how concurrence and fidelity evolves
in the presence of dephasing, for two
different values of $\Gamma_{\mathrm{deph}}$.
The panels (a) and (b) show both quantities considering
$\Gamma_{\mathrm{deph}}=1\times 10^{-4}\omega=2 \mu$eV
($0.5$ GHz), while panels (c) and (d) were obtained with $\Gamma_{\mathrm{deph}}=2\times 10^{-4}\omega=4 \mu$eV
($1$ GHz). The order of magnitude considered here for $\Gamma_\mathrm{deph}$
is in agreement with those reported on feasible experimental  scenarios~\cite{Zhang18,Shi13,Cao2013}.

From the results in Fig. \ref{fig:dephasing}, we conclude that our proposal is relatively robust against
the process of dephasing, although the entanglement degree and the fidelity present
damped oscillation (loss of coherence), the concurrence value for its
first maximum is above $C=0.6$, Figs. \ref{fig:dephasing}(a),(c), indicating a high degree of entanglement.
Analyzing the evolution of the fidelity in Fig. \ref{fig:dephasing}(b),(d), we can conclude that even when a
strong dephasing process is considered, the fidelity of the maximally entangled
state $\ket{\Psi(\pi/2)}$, given by Eq.~(\ref{eq:Bellpsi}), is up to 0.8 in its first peak.

\section{Summary}
\label{sec:summary}

We have studied a system of two qubits, encoded in two pairs of quantum dots. Inside each pair, a single electron
can tunnel between the quantum dots, thus constituting a two-level system.
Electronic degrees of freedom couple to vibrational modes.
With the aid of the Lang-Firsov transformation we show
that this coupling results in a effective electron-electron interaction,
responsible for the creation of highly entangled states.
We explore the interplay between electron tunneling,
energies detunings and the coupling between electrons and vibrational modes
on the formation of entangled states.
Our model potentially describes several experimental scenarios,
including electrons inside carbon nanotubes quantum dots or the coupling
between charged quantum dots and acoustic cavities.

Looking at the spectrum and eigenstates of the present model,
we found that, by tunning the electronic levels it is possible to
form effective two-level systems that sustain maximally entangled electronic states,
such as $\ket{\Psi(\varphi)}$ and $\ket{\Phi(\varphi)}$, as defined in Eq.(\ref{eq:Bellpsi}) and Eq.(\ref{eq:Bellphi}).
For the dynamics, we found that the electronic part of the
system can evolve to entangled states for a wide range of electron-vibrational
mode couplings. Interestingly, the frequency of the oscillations on concurrence
dynamics behaves in a non-monotonic way as the coupling parameter $g$ increases.
Using a perturbation theory that accounts for high order transition processes,
we obtain a general expression which provides the characteristic frequencies on
the dynamics, although it is valid for small values of the electron-vibrational mode coupling.

\section{Acknowledgments}
We thank the referees for helpful criticism and questions. Their careful and detailed review enriched our paper. This work was supported by CNPq (grant 307464/2015-6), and the Brazilian National Institute of Science and Technology of Quantum Information (INCT-IQ).

\appendix

\section{The two-qubit Hamiltonian in the rotated Bell-boson basis}
\label{ap1:bellboson}

The emergence of maximally entangled electronic states on the physical system of interest can be explored by writing down its Hamiltonian in terms of the electronic Bell states. Let us calculate the representation of the original Hamiltonian (\ref{eq:Hgeneral}) as a matrix written in the Bell-boson basis $\ket{\psi_{\mathrm{Bell}},ml}$, where the electronic part is ordered as \[\ket{\psi_{\mathrm{Bell}}}_{\mathcal{D}}=\{\ket{\Psi_{-}},\ket{\Phi_{-}},\ket{\Psi_{+}},\ket{\Phi_{+}}\}_{\mathcal{D}},\]. Here
\begin{eqnarray}
\ket{\Psi_{\pm}}_{\mathcal{D}}&=&\frac{1}{\sqrt{2}}\left(\ket{\uparrow\downarrow}\pm\ket{\downarrow\uparrow}\right)\nonumber\\
\ket{\Phi_{\pm}}_{\mathcal{D}}&=&\frac{1}{\sqrt{2}}\left(\ket{\uparrow\uparrow}\pm\ket{\downarrow\downarrow}\right),
\label{eq:ap1_bellstates}
\end{eqnarray}
are the well known Bell states. We choose to keep together the states with the same number of total excitations $N_{\mathcal{V}\mathrm{T}}=m+l$. This choice remarks the fact that the basis, each value of $N_{\mathcal{V}\mathrm{T}}$ defines a set of subspaces $\mathcal{S}_{\mathrm{B},(ml)}$. Notice that, for $N_{\mathcal{V}\mathrm{T}}=0$, there is only the subspace $\mathcal{S}_{\mathrm{B},(00)}$ with four states; for $N_{\mathcal{V}\mathrm{T}}=1$ we have $\mathcal{S}_{\mathrm{B},(10)}$ and $\mathcal{S}_{\mathrm{B},(01)}$ (eight states); $N_{\mathcal{V}\mathrm{T}}=3$ has twelve states associated with $\mathcal{S}_{\mathrm{B},(11)}$, $\mathcal{S}_{\mathrm{B},(20)}$, and $\mathcal{S}_{\mathrm{B},(02)}$, etc.

Let us write the matrix representation of the Hamiltonian for the first six $4$D subspaces $\mathcal{S}_{\mathrm{B},(ml)}$, ordered as $\{\mathcal{S}_{B,(00)},\mathcal{S}_{B,(01)},\mathcal{S}_{B,(10)},\mathcal{S}_{B,(11)},\mathcal{S}_{B,(02)},\mathcal{S}_{B,(20)}\}$:
\begin{equation}
H=\left(
  \begin{array}{c|cc|ccc}
    B_{00}      & G_{2}         & G_{1}         & 0             & 0             & 0             \\
  \hline
    G_{2}       & B_{01}        & 0             & G_{1}         & \sqrt{2}G_{2} & 0             \\
    G_{1}       & 0             & B_{10}        & G_{2}         & 0             & \sqrt{2}G_{1} \\
  \hline
    0           & G_{1}         & G_{2}         & B_{11}        & 0             & 0             \\
    0           & \sqrt{2}G_{2} & 0             & 0             & B_{02}        & 0             \\
    0           & 0             & \sqrt{2}G_{1} & 0             & 0             & B_{02}        \\
  \end{array}
\right).
\label{eq:ap1_HmatrixBell}
\end{equation}
By using the order $\ket{\Psi_{-},ml}$, $\ket{\Phi_{-},ml}$, $\ket{\Psi_{+},ml}$, and $\ket{\Phi_{+},ml}$, the $4$D matrices $B_{ml}$ and $ G_{v}$ are defined as
\begin{eqnarray}
\label{eq:4DBellmatrix}
B_{ml}&=&\left(
  \begin{array}{cccc}
     E_{ml} & \Delta_-   & \delta_-/2 &  0           \\
     \Delta_-   & E_{ml} &  0         & \delta_+/2   \\
     \delta_-/2 &  0         & E_{ml} & \Delta_+     \\
    0           & \delta_+/2 & \Delta_+   &  E_{ml}  \\
  \end{array}
\right),
\end{eqnarray}
and
\begin{equation}
\label{eq:4DGmatrix}
G_{v}=\left(
  \begin{array}{cccc}
   g_{v}         & 0             & 0             & 0             \\
     0           & g_{v}/2       & 0             & (-1)^{(v-1)}g_{v}/2 \\
     0           & 0             & g_{v}         & 0             \\
     0           & (-1)^{(v-1)}g_{v}/2 & 0             & g_{v}/2       \\
  \end{array}
\right),
\end{equation}
where $E_{ml}=\sum_{i=1,2}\sum_{j=3,4}\sum_{v=1,2}\left(\varepsilon_i+\varepsilon_j+\omega_v\right)$ are the energy of the state $\ket{\psi_{\mathrm{Bell}},ml}$, the tunnel couplings are defined as $\Delta_{\pm}=\Delta_{2}\pm\Delta_{1}$ and $\delta_{\pm}=\delta_{1}\pm\delta_{2}$, with $\delta_{1(2)}$ being the detuning for the qubit $1$ ($2$).

The first matrix resembles the rotated matrix on Bell basis, whose properties discussed in details on Ref.~\onlinecite{Oliveira15}, and the matrices $G_v$ depends on $g_v$ and carry on the effect of electron-vibrational mode coupling, where the factor $\sqrt{N_v}$ appears on the specific elements of the matrix (\ref{eq:ap1_HmatrixBell}) which depends on the values of $N_v$ of the coupled subspaces.

If $\delta_{\pm}=0$, it seems that the states with electronic part being $\ket{\Psi{-}}_\mathcal{D}$ and $\ket{\Phi{-}}_\mathcal{D}$ are decoupled, at the same time that $\ket{\Psi{+}}_\mathcal{D}$ and $\ket{\Phi{+}}_\mathcal{D}$ are not, in the same way that in Ref.~\onlinecite{Oliveira15}. Nevertheless, if elements for the first two lines on matrix (\ref{eq:ap1_HmatrixBell}), associated with $\ket{\Psi_-,00}$ and $\ket{\Phi_-,00}$ respectively, are written using the notation $\ket{\;}\bra{\;}$ we obtain:
\begin{widetext}
\begin{eqnarray}
\label{eq:HtermsPsi-Phi-}
H&=&\ket{\Psi_-,00}\big(E_{00}\bra{\Psi_-,00}+g_2\bra{\Psi_-,01}+g_1\bra{\Psi_-10}\big)\nonumber\\
&+&\ket{\Phi_-,00}\left(E_{00}\bra{\Phi_-,00}
+\frac{g_2}{2}\bra{\Phi_-,01}-\frac{g_2}{2}\bra{\Phi_+,01}+\frac{g_1}{2}\bra{\Phi_-,10}+\frac{g_1}{2}\bra{\Phi_+,10}\right)+...
\end{eqnarray}
\end{widetext}
 We conclude that the term with $\ket{\Psi_-}_{\mathcal{D}}$ can be written as $\big(\ket{\Psi_-}\bra{\Psi_-}\big)_{\mathcal{D}}\otimes\big(E_{00}\ket{00}\bra{00}+g_2\ket{00}\bra{01}+g_1\ket{00}\bra{10}\big)$, while the others do not permit the same.

Continuing with the calculation, we realize that only the terms on Hamiltonian associated with $\ket{\Psi_-}_{\mathcal{D}}$ are decoupled, at least from the electronic point of view, from the rest of the Bell basis. In this way, there is a Bell state, dressed by vibrational modes, becoming an eigenstate of the Hamiltonian (\ref{eq:Hgeneral}) for the specific condition of equal tunnel couplings, $\Delta_{2}=\Delta_{1}$ and the qubit detunings defined so the condition $\delta_-=0$. Writing only the terms of the Hamiltonian regarding $\ket{\Psi_{-}}_{\mathcal{D}}$ it is straightforward to see that
\begin{widetext}
\begin{eqnarray}
\label{eq:HtermsPsi-}
H_{\mathrm{with }\ket{\Psi_-}}&=&(\ket{\Psi_-}\bra{\Psi_-})_{\mathcal{D}}\otimes\big\{\big[\mathbf{\ket{00}_{\mathcal{V}}}\big(E_{00}\bra{00}_{\mathcal{V}}+g\bra{10}_{\mathcal{V}}
+g\bra{01}_{\mathcal{V}}\big)\big]
+\big[\mathbf{\ket{01}_{\mathcal{V}}}\big(E_{01}\bra{01}_{\mathcal{V}}+g\bra{11}_{\mathcal{V}}+\sqrt{2}g\bra{02}_{\mathcal{V}}\big)\nonumber\\
&&+\mathbf{\ket{10}_{\mathcal{V}}}\big(E_{10}\bra{10}_{\mathcal{V}}+g\bra{11}_{\mathcal{V}}+\sqrt{2}g\bra{20}_{\mathcal{V}}\big)\big]
+\big[\mathbf{\ket{11}_{\mathcal{V}}}\big(E_{11}\bra{11}_{\mathcal{V}}+\sqrt{2}g\bra{21}_{\mathcal{V}}+\sqrt{2}g\bra{12}_{\mathcal{V}}\big)\nonumber\\
&&+\mathbf{\ket{02}_{\mathcal{V}}}\big(E_{02}\bra{02}_{\mathcal{V}}+...\big)+\mathbf{\ket{20}_{\mathcal{V}}}\big(E_{20}\bra{20}_{\mathcal{V}}+...\big)\big]+...+\mathrm{h.c.}\big\}.
\end{eqnarray}
\end{widetext}
Other terms on Hamiltonian cannot be written as a tensorial product of the form $\ket{\psi}\bra{\psi}_{\mathcal{D}}\otimes\sum \alpha\ket{m'l'}_{\mathcal{V}}\bra{ml}$: terms with $\ket{\Psi_+}$ are coupled with $\ket{\Phi_+}$ by electron-vibrational mode interaction, while elements $\ket{\Phi_{+}}$ and $\ket{\Phi_{-}}$ are also coupled to each other by tunneling. In the Eq.~\ref{eq:HtermsPsi-}, we use bold type and the square brackets, $[\;]$, to emphasize the new Bell-boson basis $\left\{\ket{\psi_{\mathrm{Bell}},ml}\right\}$.

The number of eigenstates per ``branch", i.e. the states belonging to certain value of $N_{\mathcal{V}\mathrm{T}}$, and the number of maximally entangled electronic states at $\delta_1=0$, are connected with the dimension of original subspaces. Although these subspaces are coupled with each other, each branch can be seen as Bell-boson states, with an energy increasing as $N_{\mathcal{V}\mathrm{T}}=m+l$ grows.

\section{Numerical solutions of eigenstates of the Hamiltonian (\ref{eq:Hgeneral}) at anticrossings.}
\label{ap2:anticrossings}
In this Appendix, we present our results of the numerical calculation of the eigenstates for the Hamiltonian (\ref{eq:Hgeneral}) at each of the anticrossings discussed in Fig.\ref{fig:eigenproblem}. The results are identified by the label used along the discussion in the main text. The physical parameters used in our simulations are consistent with the same figure being $g=0.1\omega$ and $\Delta_{1}=\Delta_{2}=5\times 10^{-3}\omega$. For each case, the symbol ``$-$" (``$+$") denotes the eigenstate with lower (higher) energy from each pair on the anticrossing. For brevity, we suppressed the terms which value is less than $1\times 10^{-2}$.
\begin{widetext}
\subsection{First order anticrossings}
\begin{itemize}
\item Anticrossing A (for $\delta_2=0$):
\begin{eqnarray}
\ket{\psi_{A_-}}&\approx&-0.76\ket{\downarrow\downarrow,00}+0.58\ket{\uparrow\downarrow,00}+0.22\ket{\uparrow\uparrow,00}+0.11\ket{\downarrow\uparrow,00} +0.15\ket{\downarrow\downarrow,01}+...\\
\ket{\psi_{A_+}}&\approx&0.76\ket{\uparrow\downarrow,00}+0.58\ket{\downarrow\downarrow,00}-0.22\ket{\downarrow\uparrow,00}+0.11\ket{\uparrow\uparrow,00}-0.12\ket{\downarrow\downarrow,01}+...\nonumber
\label{eq:ap2_eigenacA}
\end{eqnarray}
\item Anticrossing B (for $\delta_2=0$):
\begin{eqnarray}
\ket{\psi_{B_-}}&\approx&-0.76\ket{\uparrow\uparrow,00}+0.58\ket{\downarrow\uparrow,00}+0.22\ket{\downarrow\downarrow,00}+0.11\ket{\uparrow\downarrow,00} +0.15\ket{\uparrow\uparrow,10}+...\\
\ket{\psi_{B_+}}&\approx&0.76\ket{\downarrow\uparrow,00}+0.58\ket{\uparrow\uparrow,00}-0.22\ket{\uparrow\downarrow,00}+0.11\ket{\downarrow\downarrow,00}-0.12\ket{\uparrow\uparrow,10}+...\nonumber
\label{eq:ap2_eigenacB}
\end{eqnarray}
\item Anticrossing A$^{\prime}$ (for $\delta_2=0.1\omega$):
\begin{eqnarray}
\ket{\psi_{A^{\prime}_-}}&\approx&0.70\ket{\uparrow\downarrow,00}-0.70\ket{\downarrow\downarrow,00}+0.14\ket{\downarrow\downarrow,01}+...
\approx\left[0.70\ket{\uparrow,00}-0.70\ket{\downarrow,00}+0.14\ket{\downarrow,01}\right]\otimes\ket{\downarrow}+...\\
\ket{\psi_{A^{\prime}_+}}&\approx&0.70\ket{\downarrow\downarrow,00}+0.70\ket{\uparrow\downarrow,00}-0.14\ket{\downarrow\downarrow,01}+...
\approx\left[0.70\ket{\downarrow,00}+0.70\ket{\uparrow,00}-0.14\ket{\downarrow,01}\right]\otimes\ket{\downarrow}+..
\nonumber
\label{eq:ap2_eigenacAprime}
\end{eqnarray}
\item Anticrossing B$^{\prime}$ (for $\delta_2=0.1\omega$):
\begin{eqnarray}
\ket{\psi_{B^{\prime}_-}}&\approx&0.70\ket{\downarrow\uparrow,00}-0.70\ket{\uparrow\uparrow,00}+0.14\ket{\uparrow\uparrow,10}+...
\approx\left[0.70\ket{\downarrow,00}-0.70\ket{\uparrow,00}+0.14\ket{\uparrow,10}\right]\otimes\ket{\uparrow}+...\\
\ket{\psi_{B^{\prime}_+}}&\approx&0.70\ket{\uparrow\uparrow,00}+0.70\ket{\downarrow\uparrow,00}-0.14\ket{\uparrow\uparrow,10}+...
\approx\left[0.70\ket{\uparrow,00}+0.70\ket{\downarrow,00}-0.14\ket{\uparrow,10}\right]\otimes\ket{\uparrow}+...\nonumber
\label{eq:ap2_eigenacBprime}
\end{eqnarray}
\end{itemize}
\subsection{Second order anticrossings}
\begin{itemize}
\item Anticrossing C (for $\delta_2=0$):
\begin{eqnarray}
\ket{\psi_{C_-}}&\approx&0.70\ket{\uparrow\downarrow,00}-0.70\ket{\downarrow\uparrow,00}\approx\ket{\Psi_-}\otimes\ket{00}+...\\
\ket{\psi_{C_+}}&\approx&-0.65\ket{\uparrow\downarrow,00}-0.64\ket{\downarrow\uparrow,00}-0.26\ket{\uparrow\uparrow,00}-0.26\ket{\downarrow\downarrow,00}+...\nonumber
\label{eq:ap2_eigenacC}
\end{eqnarray}
\item Anticrossing D (for $\delta_2=0$):
\begin{eqnarray}
\ket{\psi_{D_-}}&\approx&-0.64\ket{\uparrow\uparrow,00}-0.64\ket{\downarrow\downarrow,00}+0.27\ket{\uparrow\downarrow,00}+0.27\ket{\downarrow\uparrow,00}+0.13\ket{\uparrow\uparrow,10}+0.13\ket{\downarrow\downarrow,01}+...\nonumber\\
\ket{\psi_{D_+}}&\approx&-0.69\ket{\uparrow\uparrow,00}+0.69\ket{\downarrow\downarrow,00}+0.14\ket{\uparrow\uparrow,10}-0.14\ket{\downarrow\downarrow,01}+...
\label{eq:ap2_eigenacD}
\end{eqnarray}
\item Anticrossing C$^{\prime}$ (for $\delta_2=0.1\omega$):
\begin{eqnarray}
\ket{\psi_{C^{\prime}_-}}&\approx&-0.69\ket{\uparrow\downarrow,00}-0.69\ket{\downarrow\uparrow,00}+...\approx-\ket{\Psi_+}\otimes\ket{00}\\
\ket{\psi_{C^{\prime}_+}}&\approx&0.7\ket{\uparrow\downarrow,00}-0.7\ket{\downarrow\uparrow,00}\approx\ket{\Psi_-}\otimes\ket{00}\nonumber
\label{eq:ap2_eigenacCprime}
\end{eqnarray}
\item Anticrossing D$^{\prime}$ (for $\delta_2=0.1\omega$):
\begin{eqnarray}
\ket{\psi_{D^{\prime}_-}}&\approx&0.69\ket{\uparrow\uparrow,00}-0.69\ket{\downarrow\downarrow,00}-0.14\ket{\uparrow\uparrow,10}+0.14\ket{\downarrow\downarrow,01}+...\\
\ket{\psi_{D^{\prime}_+}}&\approx&-0.69\ket{\uparrow\uparrow,00}-0.69\ket{\downarrow\downarrow,00}+0.14\ket{\uparrow\uparrow,10}-0.14\ket{\downarrow\downarrow,01}+...\nonumber
\label{eq:ap2_eigenacDprime}
\end{eqnarray}
\end{itemize}
\end{widetext}
\end{document}